\documentclass{PoS}
\usepackage{amsmath}
\usepackage{tabularx}
\usepackage{graphicx}
\usepackage{longtable}
\usepackage{caption}
\usepackage{url}

\title{Localization in SU(3) gauge theory}

\ShortTitle{Localization transition}

\author{\speaker{R\'eka \'A. Vig}\thanks{Supported by OTKA Hungarian Science Fund under Grant No. OTKA-K-113034.
}\\
       University of Debrecen, Hungary\\
        E-mail: \email{vig.reka@atomki.mta.hu}}

\author{Tam\'as G. Kov\'acs\footnotemark[2]\\
       Institute for Nuclear Research, Debrecen, Hungary\\
        E-mail: \email{kgt@atomki.mta.hu}}

\abstract{In this paper we study the localization transition of Dirac eigenmodes in quenched QCD. We determined the temperature dependence of the mobility edge in the quark-gluon plasma phase near the deconfining critical temperature. We calculated the critical temperature where all of the localized modes disappear from the spectrum and compared it with the critical temperature of the deconfining transition. We found that the localization transition happens at the same temperature as the deconfining transition which indicates a strong relation between the two phenomena.
         }

\FullConference{The 36th Annual International Symposium on Lattice Field Theory - LATTICE2018\\
		22-28 July, 2018\\
		Michigan State University, East Lansing, Michigan, USA.}

\begin{document}

\section{Introduction}
In QCD there is a crossover at high temperatures where the quarks go from a bound hadronic state to a quasi free quark gluon plasma (QGP) state.
 Around the temperature of the crossover the properties of strongly interacting matter change rapidly and besides the deconfining transition of the quarks two other phenomena occur. 
On the one hand while the approximate chiral symmetry is spontaneously broken at low temperatures, it becomes restored in the high temperature regime. 
This is accompanied by the localization transition of the lowest end of the Dirac spectrum, which means that the low energy eigenmodes become localized as we go to high temperatures. 
In contrast it is known that at lower temperatures all modes of the Dirac operator are extended.
Since these phenomena happen during the crossover, the question arises whether they are in a causal connection. 
For the localization of the low modes we can assign an exact temperature where the localized modes appear,
 whereas the other phenomena are just crossovers so we cannot tell wether they happen exactly at the same temperature as the localization transition. 
However, there are models where a genuine deconfining or chiral phase transition occurs, for example in quenched QCD. So in this case we can ask wether the critical temperatures of the different transitions are equal. If they happen at the same temperature, that would mean there is a strong connection between the three phenomena.

In this paper we will use the SU(3) quenched theory with a genuine first order deconfining phase transition and study the localization transition and how it is connected to the deconfining phase transition.
 We simulated SU(3) lattices above the deconfining phase transition and used lattice configurations of different temperatures (or different $\beta$ gauge couplings) near the critical temperature. 
We wanted to see how the localised modes behave as we go closer to the deconfining critical temperature.
 For this we calculated the low end of the spectrum of the Dirac operator for each configuration and determined the so called 'mobility edge' \cite{Kovacs:2012zq}, the energy value which separates the localized modes from the extended  modes in the spectrum. 
After this  we could analyze how this value tends toward zero as we decrease the temperature.
 We cannot see this directly from the lattice simulations, because at the critical temperature the correlation length is too large and this is not beneficial computationally to use that large lattices.
 So we extrapoleted where the mobility edge reaches zero which is the critical temperature of the localization transition.
Below that temperature there are no localized modes, all of the modes are extended in the spectrum.
 We used the spectra of staggered fermions and the overlap operator \cite{Narayanan:1993sk} and found that for both cases the critical temperature of the localization coincides with the critical temperature of the deconfining phase transition which we know fom the literature for the SU(3) quenched gauge theory \cite{Francis:2015lha}.
 This means that there is a strong connection between the deconfining of the quarks and the localization of the low energy modes.
Results by Holicki et al. presented at this conference show that twisted-mass Wilson quarks also exhibit a localization transition \cite{Holicki:2018sms}. Another interesting question is how the localized modes related to instanton zero modes. Which is discussed at this conference by T. Kovács.

In the next section we will see the method we used to calculate the mobility edge $\lambda_c$ and after that in Section 3 we will show how we extrapolated the $\lambda_c(T)$ function to get the results for the critical temperature of the localization transition.

\section{Determination of the mobility edge}
Our aim is to find the critical temperature where the localized modes of the Dirac operator completely disappear. For this we study the low part of the Dirac spectrum at high temperatures just above the deconfining critical temperature. We know that at high temperatures the lowest end of the spectrum of the Dirac operator consists of localized modes while higher in the spectrum all modes are extended. The mobility edge $\lambda_c$ is the  boundary which separates the localized modes from the extended modes. As we decrease the temperature the mobility edge moves down in the spectrum toward zero and at a critical temperature $T_c^{loc}$ it reaches zero $\lambda_c(T_c^{loc})=0$. At this point all of the eigenmodes become extended even in the lowest part of the spectrum. We want to find this transition in the quenched theory where we set the temperature by the inverse gauge coupling $\beta$ of the Wilson gauge action. We calculate the mobility edge as the function of the inverse gauge coupling and extrapolate it to get the critical coupling $\beta_c^{loc}$ of the localization transition and compare it with the known value of the deconfining transition $\beta_c^{deconf}$.

Now we will show the method we used to calculate the mobility edge.
To find the energy which separates the localized modes from the extended modes first we have to analyze the local statistical properties of the spectrum. 
\begin{figure}
\begin{center}
\includegraphics[width=0.95\columnwidth,keepaspectratio]{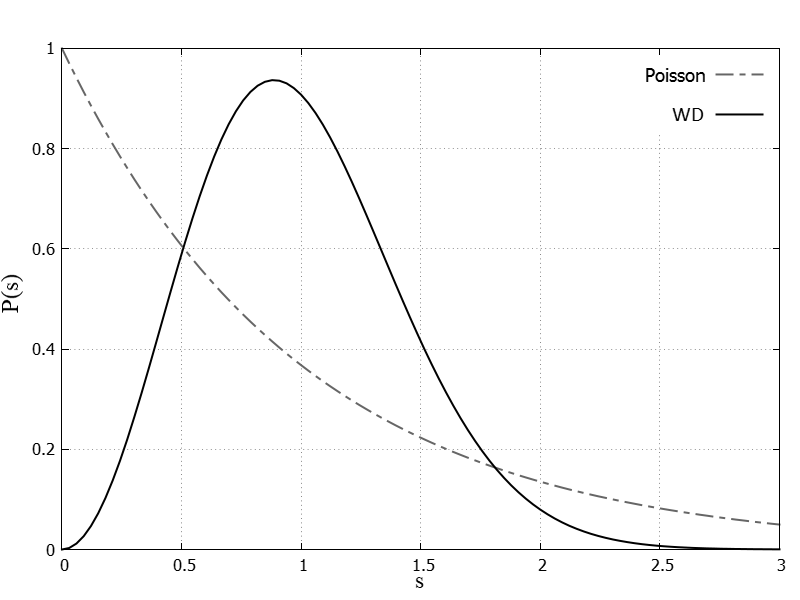}
\caption{ \small The level spacing probability density functions. The dashed line represents the Poisson distribution, the solid line is for the Wigner-Dyson distribution.}
\end{center}
\end{figure}
 The localized modes are independent of each other therefore the corresponding eigenvalues will be described by Poisson statistics. 
The extended modes are mixed by the gauge field therefore the eigenvalues of this part of the spectrum will obey Wigner-Dyson statistics which we know from random matrix theory \cite{universality}. 
We studied a quantity which is known analitically for both kinds of statistics, the so called unfolded level spacing distribution. This is the probability distribution of the energy differences between two adjacent eigenvalues of the Dirac operator. Unfolding means a local rescaling of the level spacings by their average value. By this we set the spectral density to unity through the whole spectrum to get a universal, scale independent quantity.  
In Figure 1 we show the probability density functions of the level spacings for the two kinds of statistics.

\begin{figure}
\begin{center}
\includegraphics[width=0.95\columnwidth,keepaspectratio]{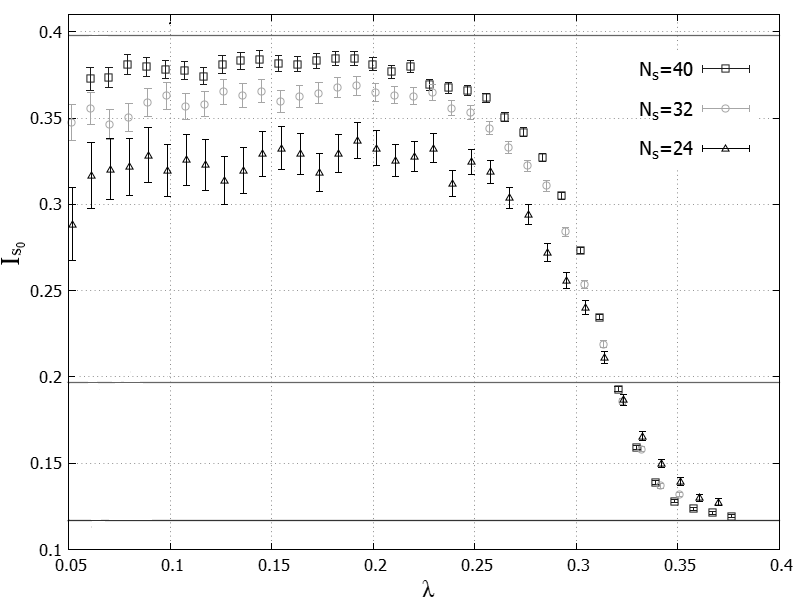}
\caption{\small The function $I_{s_0}(\lambda)$ along the spectrum ($\lambda$) for staggered fermions, $N_t=4, \beta=5.74$ with
  three different spatial volumes. }
\end{center}
\end{figure}

 We can detect the transition in the spectrum where the statistics changes from Poisson to Wigner-Dyson
by calculating a parameter of the level spacing distribution and tracing how it changes as we move along in the spectrum. 
The parameter we choose is the integrated probability density function of the unfolded level spacing distribution
\begin{align}
& I_{s_0}= \int_{0}^{s_0}p(s)ds.
\end{align}
Here $s$ is the difference between two adjacent eigenvalues, $s_{0}$ is a parameter and $p(s)$ is the probability density function of the unfolded level spacings. It is $p(s) =\exp(-s)$ in the case of localized modes. In the case of extended modes it is $p(s) = \dfrac{32}{\pi^{2}}s^{2}\cdot
                       \exp\left(-\dfrac{4}{\pi}s^{2}\right)
$, the unitary Wigner surmise \cite{universality} which belongs to the same universality class as our Dirac operator.
 We chose the first intersection point of the two density functions for the upper bound of the integral $s_{0}=0.508$, because in this case the difference between the integrals of the two limiting statistics is maximized. To trace the change of this quantity we divide the spectrum into small intervals and calculate $I_{s_0}$ in each of them separately.

In Figure 2 we show a typical graph of how $I_{s_0}$ changes in the spectrum for three different spatial volumes. 
The transition becomes sharper as we go to larger volumes.
It was shown that in the infinite volume limit this becomes a genuine second order Anderson type transition belonging to the three dimensional unitary class \cite{Giordano:2013taa}.      
The critical point of this transition in the spectrum is the mobility edge $\lambda_c$. 
Since we work with finite volume lattices the transtion is not infinitely sharp as it is in the thermodynamic limit.
We can define $\lambda_c$ to be the point where the function $I_{s_0}$  reaches an arbitrarily chosen value between the Poisson and Wigner-Dyson values. In the infinite volume limit $\lambda_c$ will be exact, since any value chosen between the limiting $I_{s_0}^{Poisson}=0.398$ and $I_{s_0}^{WD}=0.117$ values tend to the same point in the spectrum as we move to infinite volumes.

\LTcapwidth=\textwidth
\begin{longtable}{|p{0.3cm}p{0.8cm}rrr|p{0.3cm}p{0.6cm}rrr|p{0.3cm}p{0.6cm}rrr|}
 \multicolumn{15}{c}{Staggered fermions} \\ \hline
 $N_t $ & $\beta$ & $N_s$ & Nconf & Nevs & $N_t $ & $\beta$ & $N_s$ & Nconf & Nevs &  $N_t $ & $\beta$ & $N_s$ & Nconf & Nevs \\ \hline 
      4 & 5.693  & 32    & 1061   &  600  &  6 & 5.897    & 32    &  532   &  900 &8 & 6.1    & 48    &  439   & 500  \\
        &        & 40    & 1192   &  1100 & &        & 40    &  1249   &  1000 &  &        & 56    &  681   & 600 \\ 
        &        & 48    & 2381   &  1500&&        & 48    & 682   & 1350&  &        & 64    &  486   & 400 \\ 
        & 5.694  & 32    &  1715   &  600 && 5.9    & 32    &  998   &  900&  & 6.15   & 48    &  781   & 500\\
        &        & 40    & 1005   &  1100&&        & 40    &  925   &  1000  &  &        & 56    &  698   & 600\\ 
        &        & 48    & 2014   &  1600 &&        & 48    & 813   & 1350&&        & 64    &  385   & 400\\
        & 5.695 & 32    & 2184    &  650 && 5.91   &32    & 1068   & 900&& 6.18   & 48    &  636   & 500 \\
        &        & 40    & 2012   & 1100 &&        & 40    & 834   & 1000 &&        & 56    &  964   & 600\\ 
        &        & 48    &  2028   & 1300 & &        & 48    & 1088   & 1350&   &        & 64    &  384   & 400  \\ 
        & 5.696 & 32    & 1073    & 900  & & 5.92   & 32    & 1822   & 600& & 6.2    & 48    &  675   & 500\\ 
        &        & 40    & 1628   &  1000 &&        & 40    & 960   & 1000& &        & 56    &  778   & 600\\ 
        & 5.6975 & 32    & 2291    & 600&  & 5.93   & 32    & 806   & 900 &&        & 64    &  418   & 400 \\ 
        &        & 40    & 1524   & 1100 & &        & 40    & 1050   & 1000 && 6.25   & 48    &  758   & 500 \\ 
        &        & 48    &  2000   &  1500 &  & 5.94   & 40    & 1092   & 1000& &        & 56    &  652   & 600\\ 
        & 5.6985 & 40    &  1973   &  1000 && 5.95   & 32    & 562   & 600&&        & 64    &  320   & 400 \\
        & 5.7 & 24    &  4139   & 600 & &        & 40    & 1276   & 1000 && 6.3    & 48    &  578   & 500\\
        &        & 32    & 4040   &  800 && 5.96   & 32    & 832   & 1000& &        & 56    &  616   & 600\\
        &        & 40    & 1022   & 1000  &  &        & 40    & 1032   & 1000&&        & 64    &  452   & 400 \\
        & 5.71 & 24    &  2509   &  300 & & 6.0   & 32    & 1392   & 900& &&&&\\ 
        &        & 32    & 2507   &  450 & &        & 40    & 1958   & 1000&&&&&\\ 
        &        & 40    &  1073   &  1100& &&&&&&&&&\\ 
        & 5.74 & 24    &  2024   &  300 &&&&&&&&&&\\ 
        &        & 32    & 2501   &  450 &&&&&&&&&&\\ 
        &        & 40    &  2390   &  1100& &&&&&&&&&\\\hline 
 \multicolumn{15}{c}{ }\\ 
 \multicolumn{15}{c}{Overlap fermions} \\ \cline{6-10}
\multicolumn{5}{c|}{ }& $N_t $ & $\beta$ & $N_s$ & Nconf & Nevs &\multicolumn{4}{|c}{ } \\\cline{6-10}
\multicolumn{5}{c|}{ } & 6 & 5.91    & 40    &  741   &  80&\multicolumn{4}{|c}{ } \\
\multicolumn{5}{c|}{ }    & & 5.92    & 40    &  797   &  80&\multicolumn{4}{|c}{ }\\
\multicolumn{5}{c|}{ }     & &5.93   &40   & 750   & 80&\multicolumn{4}{|c}{ } \\
\multicolumn{5}{c|}{ }       & & 5.94   & 40    & 770  & 80 &\multicolumn{4}{|c}{ }\\
\multicolumn{5}{c|}{ }        & &5.95   & 40    & 565   & 80&\multicolumn{4}{|c}{ } \\
\multicolumn{5}{c|}{ }        & &5.96   & 40    & 605   & 80&\multicolumn{4}{|c}{ } \\\cline{6-10}
\caption{ \small The parameters of the simulations. $N_t$ is the
   temporal size of the lattices, $\beta$ is the inverse Wilson gauge coupling, the $N_s$ is 
   spatial size of the lattices, Nconf is the number of configurations and Nevs is the number of
 eigenvalues computed on one configuration. The parameters for the staggered operator are up and for the overlap operator are below.}
\end{longtable} 
 \noindent  We used the value $ I^{crit}_{s_0}=0.1966$, which was obtained in a finite size scaling study \cite{Giordano:2013taa} , to locate $\lambda_c$ in the spectrum because in this case finite size corrections are small. To determine $\lambda_{c}$ from the data we used a linear fit to find where the function reaches its  $ I^{crit}_{s_0}$ critical value. We can do this because around  $ I^{crit}_{s_0}$ the function has an inflection point therefore we could approximate it with a linear function on a small interval. We increased the spatial volumes by steps and calculated the mobility edge for each of them up to the point where $\lambda_c$ did not change anymore within the statistical error and used the last volumes for the calculations. Colse to the deconfining critical teperature we needed to use larger spatial volumes because of the increasing correlation length. The error was calculated by the Jackknife method. In Table 1 we show the parameters of the simulations. We simulated quenched SU(3) gauge theory with the Wilson gauge action. The temporal extent of the lattices were $N_t=4, 6, 8$ for staggered fermions and $N_t=6$ for the overlap operator with two stout smearings \cite{Aoki:2005vt} in both cases. We calculated the mobility edge for different gauge couplings above the deconfining critical temperature.

\section{Results for the critical temperature of the localization transition}
\begin{figure}
\begin{center}
\includegraphics[width=0.95\columnwidth,keepaspectratio]{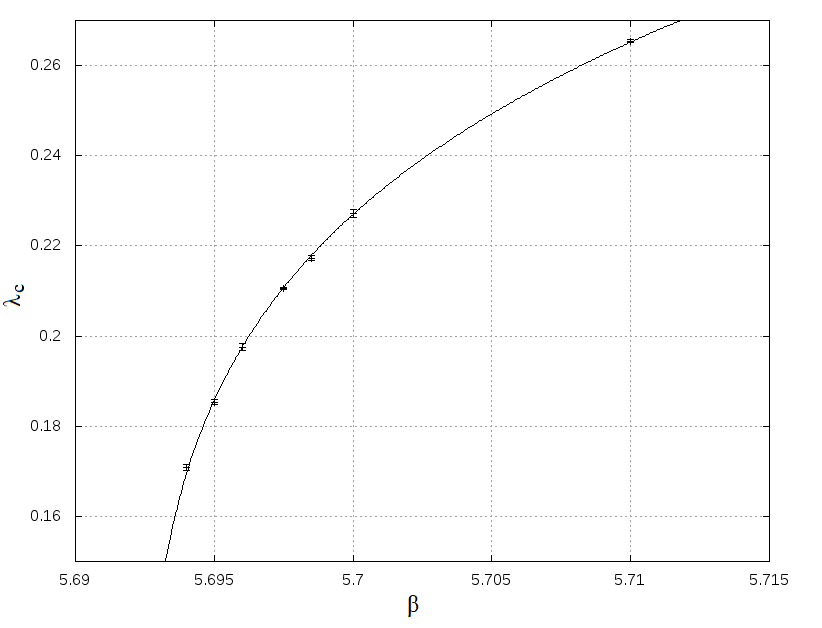}
\caption{ \small The mobility edge as a function of the inverse gauge coupling for staggered fermions and temporal extent $N_t=4$  with a fitted function.}
\end{center}
\end{figure}
Now we will show how we determined the critical coupling of the localization transition. For this we have to analyze the dependence of the mobility edge on the inverse gauge coupling $\beta$. We varied $\beta$ above the deconfining critical temperature and for each $\beta$ calculated the mobility edge as shown in the  previous Section. During this procedure the temporal size $N_t$ of the lattices was kept fixed. 
Since the correlation length grows as we go toward the transition, to keep up with that we needed to use lattices with larger spatial volume. We used the largest spatial volume we could afford which made a restriction on the smallest $\beta$ we could use. We fitted a three parameter function to our data: 
\begin{align}
 \lambda_c(\beta)=p_1(\beta-\beta_c^{loc})^{p_2}.
\label{eq:myeqn}
\end{align}
Here $ p_1, p_2$ and $\beta_c^{loc}$ are the parameters. We found that this function describes the data well in a range above the critical temperature. We limited the upper end of the fit range by requiring an acceptable value for $\chi^2$.
The critical coupling is the parameter $\beta_c^{loc}$ where the mobility edge reaches zero $\lambda_c(\beta_c^{loc})=0$  and all of the localized modes disappear from the Dirac spectrum.
In Figure 3 we show data for $N_t=4$ with a fitted function $\lambda_c(\beta)$.  In the other cases with different temporal sizes the graphs look very similar.
\bgroup
\def\arraystretch{1}
\begin{table}[htb]
\centering
\begin{tabular}{lllllll}
\hline
\hline
& $N_t$ &\centering $\beta_{c}$\textsuperscript{deconf} &
\centering$\beta_{c}$\textsuperscript{loc} & $p_1$ & $p_2$ & fit range
   \tabularnewline   \hline
Staggered &4 & 5.69254(24)  & 5.69245(17)  & 0.1861(6) & 0.563(2)  & 5.695-5.71 \\
&6 & 5.8941(5)    & 5.8935(16)   & 0.1580(8) & 0.320(1)  & 5.91-5.96  \\ 
&8 & 6.0624(10)   & 6.057(4)     & 0.164(4)  & 0.233(2)  & 6.08-6.18  \\ 
\hline
\hline     
Overlap & 6 &5.8941(5)&5.8943(85)& 0.170(2) &0.200(4)&5.91-5.96\\
\hline
\hline
 \end{tabular}
\caption{ \small Main results for the critical couplings with staggered and overlap fermions. The $N_t$ is the temporal size of the lattice, $\beta_c$\textsuperscript{deconf} and $\beta_c$\textsuperscript{loc} are the critical inverse gauge couplings for the deconfining and the localization transition, $p_1$ and $p_2$ are the fit parameters and we also show the range of the fit we used.}
\end{table}
\egroup
 
Previously we used this procedure for temporal extents of $N_t=4,6$ and $8$ with staggerred fermions \cite{Kovacs:2017uiz} and found that within the statistical error the critical couplings of the localization transition coincide with the critical couplings of the deconfining transition.
 Subsequently we also calculated the critical coupling for $N_t=6$ with overlap fermions where we used the same lattice ensemble as in the staggered case. It also showed good agreement with the critical coupling of the deconfining transition. In Table 2 we show the quantitative results in the four cases.

\section{Summary}
We studied the localization transition of the quarks in quenched SU(3) lattice gauge theory. We calculated the energy that separates the localized modes from the extended modes, called the mobility edge. The temperature was set by the inverse gauge coupling. By changing the inverse gauge coupling we determined the temperature dependence of the mobility edge and extrapolated it to find the critical temperature of the localizaion transition. The critical point is where the mobility edge reaches zero, therefore here the localized modes disappear and under that temperature all of the modes are extended. We calculated the critical temperature for both staggered and overlap fermions and found good agreement with the critical temperature of the deconfining transition. This indicates that these two phenomena are strongly related.


\begin{thebibliography}{99}
\bibitem{Kovacs:2012zq}  T.~G.~Kov\'acs, F.~Pittler, \textit{ Poisson to random matrix transition in the QCD Dirac spectrum}, Phys.\ Rev.\ D {\bf 86}, 114515 (2012).
\bibitem{Narayanan:1993sk} 
  R.~Narayanan and H.~Neuberger,
  \textit{Chiral determinant as an overlap of two vacua},
  Nucl.\ Phys.\ B {\bf 412}, 574 (1994).
  R.~Narayanan and H.~Neuberger,
  \textit{A Construction of lattice chiral gauge theories},
  Nucl.\ Phys.\ B {\bf 443}, 305 (1995).
\bibitem{Francis:2015lha} A.~Francis, O.~Kaczmarek, M.~Laine, T.~Neuhaus, and H.~Ohno, \textit{Critical point and scale setting in SU(3) plasma: An update}, Phys.\ Rev.\ D {\bf 91}, no. 9, 096002 (2015).
\bibitem{Holicki:2018sms}Lukas~Holicki, Ernst-Michael~Ilgenfritz, Lorenz~von~Smekal, \textit{The Anderson transition in QCD with Nf = 2+1+1 twisted mass quarks: overlap analysis}, arXiv:1810.01130 [hep-lat].
\bibitem{universality}J.~J.~M.~Verbaarschot, T.~Wettig, \textit{Random Matrix Theory and Chiral Symmetry in QCD}, Ann.\ Rev.\ Nucl.\ Part.\ Sci.\  {\bf 50}, 343 (2000).
\bibitem{Giordano:2013taa}  Matteo~Giordano, Tam\'as~G.~Kov\'acs, Ferenc~Pittler, \textit{Universality and the QCD Anderson Transition}, Phys.\ Rev.\ Lett. \ {\bf 112},102002 (2014).
\bibitem{Aoki:2005vt} 
  Y.~Aoki, Z.~Fodor, S.~D.~Katz and K.~K.~Szabo,
 \textit{The Equation of state in lattice QCD: With physical quark masses towards the continuum limit}, JHEP {\bf 0601}, 089 (2006).
\bibitem{Kovacs:2017uiz} Tam\'as~G.~Kov\'acs, R\'eka~\'A.~Vig, \textit{Localization transition in SU(3) gauge theory}, Phys.\ Rev.\ D {\bf 97}, no. 1, 014502 (2018).



\end{thebibliography}
\end{document}